\documentclass[aps,superscriptaddress,eqsecnum,nofootinbib,showpacs,preprintnumbers]{revtex4-2}
\usepackage[utf8]{inputenc}
\usepackage{float}
\usepackage{hyperref}
\usepackage{amsmath}
\usepackage{amsfonts}
\usepackage{latexsym}
\usepackage{xcolor}
\usepackage{graphicx}
\usepackage{xcolor}
\usepackage{framed}
\usepackage{amssymb}
\usepackage{graphicx, rotating}
\usepackage{epsfig}
\usepackage{latexsym}
\usepackage{bm}
\usepackage{slashed}
\usepackage{hyperref}
\usepackage{epstopdf}
\usepackage{slashed}
\usepackage{tabularx}

\begin{document}
\title{Thermodynamic Phase Transitions in Einstein-Maxwell-Scalar-Gauss-Bonnet Gravity}
\author{Cristi\'an \surname{Erices}}
\email{cristian.erices@ucentral.cl}
\author{Stella Kiorpelidi}
\email{stellak.postdoc@ucentral.cl}
\affiliation{Centro de Investigación en Ciencias del Espacio y Física Teórica
(CICEF), Universidad Central de Chile, La Serena 1710164, Chile}
\affiliation{Physics Department, National Technical University of Athens, 15780 Zografou Campus, Athens, Greece}

\begin{abstract}
Although asymptotically flat black holes generically lack thermodynamic phase transitions, we show that curvature-induced scalarization of electrically charged black holes in Einstein-Maxwell-scalar-Gauss-Bonnet theory provides a natural setting for nontrivial thermodynamic behavior, without invoking external confining mechanisms or an extended thermodynamic formalism. Working within the canonical ensemble and employing the Euclidean approach, we identify the coexistence of locally stable scalarized and Reissner-Nordstr\"{o}m thermal states, which promotes free-energy crossings to bona fide phase transitions between equilibrium phases. For weak coupling, a second-order phase transition coincides with the second bifurcation point, at which the scalarized branch reconnects with the Reissner-Nordstr\"{o}m branch and scalar hair is spontaneously shed. As the coupling strength increases, this transition becomes zeroth order, the scalarized branch shrinks, and a fish-like structure develops in its Helmholtz free energy, rendering locally stable thermal states partially metastable, and yielding up to three phase transitions. In the strong-coupling limit, the scalarized branch reduces to a Schwarzschild-like solution, and the Reissner-Nordstr\"{o}m phase ultimately emerges as the sole thermodynamically preferred configuration.
\end{abstract}

\maketitle

\section{Introduction}

Black hole thermodynamics provides a profound link between gravity, quantum theory, and statistical mechanics \cite{Carlip:2014pma}. The pioneering works of Bekenstein \cite{Bekenstein:1973ur} and Hawking \cite{Hawking:1975vcx} established that black holes can be consistently regarded as thermodynamic systems endowed with entropy and temperature. This picture was further strengthened by Bardeen, Carter, and Hawking, who showed that stationary black holes obey four laws of black hole mechanics which closely parallel the laws of ordinary thermodynamics \cite{Bardeen:1973gs}. A decisive step toward a statistical interpretation of black hole thermodynamics was provided by the Euclidean approach of Gibbons and Hawking \cite{Gibbons:1976ue,Hawking:1976de}, in which the gravitational partition function is defined through a path integral over smooth Euclidean geometries with periodic imaginary time. In the semiclassical limit, the path integral is dominated by classical Euclidean solutions, and the corresponding on-shell Euclidean action plays the role of a thermodynamic potential.

A large body of subsequent work has investigated black hole spacetimes as thermodynamic systems defined within different statistical ensembles, depending on which macroscopic quantities are held fixed at the boundary. In general relativity (GR), stationary black hole solutions, namely the Schwarzschild, Reissner-Nordstr\"{o}m (RN), and Kerr solutions, are completely characterized by a small set of global parameters: mass, electric charge, and angular momentum, according to the no-hair theorem \cite{Ruffini:1971bza}. The corresponding thermodynamic state of a black hole is then described by extensive variables such as the entropy, electric charge, and angular momentum, together with their conjugate intensive quantities, namely the Hawking temperature, the electric potential, and the angular velocity, respectively.

A central role in the thermodynamic analysis is played by the response functions of the system, in particular the specific heats, whose sign determines the local thermodynamic stability of a given configuration under thermal fluctuations. For a Schwarzschild black hole, the specific heat is negative, reflecting the well-known fact that it becomes hotter as it loses energy through Hawking radiation and is therefore locally thermodynamically unstable \cite{Hawking:1974rv}. For a Reissner-Nordstr\"{o}m black hole at fixed electric charge, the specific heat diverges at a particular mass. This divergence does not correspond to a phase transition in ordinary black hole thermodynamics, but instead signals a change in the sign of the specific heat, separating locally unstable and locally stable branches. This point is commonly referred to as the Davies point \cite{Davies:1978zz}. The Kerr black hole exhibits an analogous qualitative behavior.

Global thermodynamic stability is determined by the relevant thermodynamic potential, whose minimum selects the preferred equilibrium configuration of the system. A phase transition occurs when two distinct states exchange global dominance of this potential. The first phase transition, identified by Hawking and Page \cite{Hawking:1982dh} for asymptotically anti-de Sitter (AdS) spacetimes, is a first-order transition between thermal AdS space and a large Schwarzschild-AdS black hole. It is known as the Hawking-Page phase transition and is commonly interpreted as the gravitational analog of a solid-liquid phase transition \cite{Kubiznak:2014zwa}. Beyond this prototypical example, charged AdS black holes display a small black hole/large black hole phase transition that is closely analogous to the liquid-gas transition of a van der Waals fluid and is characterized by the familiar swallowtail structures in the thermodynamic potentials \cite{Chamblin:1999tk}. 

An even broader class of thermodynamic phenomena arises within the framework of extended black hole thermodynamics, where the negative cosmological constant is interpreted as a thermodynamic pressure \cite{Kubiznak:2016qmn}. Within this perspective, a wide variety of phase structures has been uncovered, closely analogous to those found in conventional chemical and condensed matter systems. These include re-entrant phase transitions \cite{Altamirano:2013ane}, triple points \cite{Altamirano:2013uqa,Wei:2014hba}, and other novel critical phenomena, particularly in the context of modified theories of gravity (see, e.g., \cite{Mann:2025xrb} for a recent review).

Despite the remarkable richness of phase phenomena uncovered in asymptotically AdS spacetimes, the thermodynamics of asymptotically flat black holes remains comparatively constrained within general relativity. This limitation is closely related to the fact that, in GR, asymptotically flat black holes possess only a limited set of global charges and lack a natural confining scale, such as the AdS radius. Consequently, their thermodynamic properties are predominantly determined by the universal features associated with Hawking radiation and their negative specific heat \cite{Hawking:1982dh}. Therefore, thermodynamically stable configurations and nontrivial phase structures are difficult to realize without the introduction of external regulators, such as a finite cavity \cite{York:1986it,Lundgren:2006kt}, or considering extended thermodynamic frameworks \cite{Astefanesei:2019ehu}.

In a different setting, scalar-tensor theories of gravity extend general relativity by introducing one or more dynamical scalar fields that couple to curvature invariants and/or to matter fields in addition to the spacetime metric. These couplings give rise to distinct classes of hairy black hole solutions that evade the standard no-hair theorems \cite{LuHui,Herdeiro:2015waa,Yazadjiev:2025ezx}. Such theories are well-motivated, as scalar fields are ubiquitous in low-energy effective field theories \cite{Burgess:2007pt} and naturally arise in ultraviolet completions such as string theory \cite{Grana:2005jc,Polchinski:1998rq}.

Nontrivial scalar configurations are known to substantially alter the thermodynamic properties and local stability of black holes in AdS asymptotics \cite{Bravo-Gaete:2022mnr,Caceres:2023gfa,Erices:2024iah}, and even in asymptotically flat spacetimes \cite{Astefanesei:2019qsg}. For example, in theories with a self-interacting dilaton field, charged black hole solutions have been shown to remain thermodynamically stable under mixed thermal and electric fluctuations \cite{Astefanesei:2009wi,Astefanesei:2019mds,Astefanesei:2020xvn,Astefanesei:2024wfj}. In shift-symmetric Beyond Horndeski theories, primary hair enhances local stability relative to the Schwarzschild solution \cite{Erices:2024lci}, and extends the first law of black hole thermodynamics by introducing an additional work term associated with a conserved scalar charge that emerges naturally from the shift symmetry \cite{Bakopoulos:2024ogt}.

In this context, a particularly interesting scenario is realized in the strong gravity regime, where higher-order curvature corrections to general relativity become significant, yet their inclusion generically introduces ghost instabilities \cite{stelle}. One way to circumvent this difficulty is provided by the Gauss-Bonnet invariant: although topological, and therefore dynamically trivial in four dimensions, it acquires propagating degrees of freedom when nonminimally coupled to a scalar field, and does so free of Ostrogradsky instabilities. This construction defines Einstein-scalar-Gauss-Bonnet (EsGB) gravity, a framework that has attracted sustained research interest in recent years, particularly regarding hairy black hole solutions and compact objects in four-dimensional spacetimes \cite{stewart,kanti,torii2,yunes,radu,kleihaus}, and even in cosmological scenarios\cite{Erices:2022bws}. 

A remarkable feature of this class of theories is the phenomenon of spontaneous scalarization, a distinctive manifestation of gravitational interactions in the strong-field regime first identified three decades ago by Damour and Esposito-Far\`ese \cite{DEF} in the context of neutron stars, finding non-perturbative deviations from GR. This phenomenon was soon recognized to resemble a continuous second-order phase transition describable by a Landau-type phenomenological model \cite{Damour:1996ke}, in close analogy with spontaneous magnetization in ferromagnets. It was later discovered that scalarization can also occur in black holes through nonminimal couplings between a scalar field and higher-curvature invariants or matter fields \cite{Doneva:2017bvd,Silva:2017uqg,Herdeiro:2018wub}. In this scenario, the field equations admit both a trivial general relativistic branch and a scalarized branch with a nontrivial scalar profile; below a critical mass threshold, the former develops a tachyonic instability in regions of strong curvature, and the scalarized branch emerges continuously as the dynamically favored outcome, displaying distinct physical and thermodynamic properties. This has motivated the systematic exploration of diverse coupling functions and generalizations of the scalarization phenomenon \cite{Doneva:2018rou,herdeiro2,Dima:2020yac}, as reviewed comprehensively in \cite{Doneva:2022ewd} and references therein.

While most studies have focused on the existence, dynamical stability, and phenomenological properties of scalarized configurations, a systematic thermodynamic analysis has only recently begun to receive attention \cite{Unluturk:2025zie, Staykov:2026ojk}. A particular step in this direction was taken in \cite{Herdeiro:2026sur}, where the phase structure of neutral scalarized black holes in EsGB theory was investigated within a Landau-type framework, with the scalar charge playing the role of an order parameter. A central outcome of that analysis is that the phase structure is remarkably sensitive to the coupling parameter $\beta$: depending on its value, the transition between the Schwarzschild and the scalarized branch may be of second order, of first order, or absent altogether. In this framework the scalarized branch bifurcates from the Schwarzschild solution at the onset of the instability and can be either locally unstable in the weak-coupling regime or locally stable in the strong-coupling regime, even though the Schwarzschild branch itself is locally thermodynamically unstable due to its negative specific heat. The presence of an electric charge is expected to qualitatively enrich this picture: unlike their neutral counterparts, small Reissner-Nordstr\"{o}m black holes possess positive specific heat and are therefore locally thermodynamically stable, so that the sensitivity to $\beta$ now unfolds against a locally stable trivial phase, opening the door to novel phase transitions between coexisting equilibrium configurations.

In this work, we investigate precisely this setting by analyzing the thermodynamic phase structure of Reissner-Nordstr\"{o}m and curvature-induced scalarized black holes in the canonical ensemble, within Einstein-Maxwell-scalar-Gauss-Bonnet (EMsGB) theory, where the scalar field is nonminimally coupled to the Gauss-Bonnet invariant in the presence of an electromagnetic field \cite{Doneva:2018rou}. We show that this setup provides a natural framework in which nontrivial thermodynamic phenomena arise for asymptotically flat black holes without invoking external confining mechanisms or an extended thermodynamic framework.

This paper is organized as follows. In Sec.~\ref{thescalarization}, we briefly review the scalarization mechanism in EMsGB gravity and summarize the main properties of the corresponding black hole solutions. In Sec.~\ref{euclidean thermodynamics}, we employ the Euclidean approach to define the associated thermal states and construct the relevant thermodynamic potentials. Sec.~\ref{Thermodynamical analysis} is devoted to the local and global thermodynamic stability analysis of the Reissner-Nordstr\"om and scalarized branches, while Sec.~\ref{phase structure} presents the global thermodynamic phase structure, which exhibits features that can be broadly organized into three coupling regimes, each admitting a further sub-structure. Finally, Sec.~\ref{conclusions} summarizes our main results and discusses their physical implications.

In this work, we set $G=c=1$.

\section{Curvature-induced scalarized black holes} \label{thescalarization}
The action describing the EMsGB theory is given by:
\begin{equation}
    S=\frac{1}{16\pi} \int\,\mathrm{d}^4x\,\sqrt{-g}\left(R-2\nabla_\mu\phi\nabla^\mu\phi+\lambda^2f(\phi)\mathcal{R}_{GB}^2-F_{\mu\nu}F^{\mu\nu}\right),\label{action}
\end{equation}
where $F_{\mu\nu}$ is the Maxwell tensor, defined as $F_{\mu\nu}=\nabla_\mu A_\nu-\nabla_\nu A_\mu$, and $A_\mu=\left(V(r),0,0,0\right)$ is the electromagnetic four-potential in the presence of an electric field. The Gauss-Bonnet invariant $\mathcal{R}_{GB}^2$ is defined as $\mathcal{R}_{GB}^2=R^2-4R_{\mu\nu}R^{\mu\nu}+R_{\mu\nu\alpha\beta}R^{\mu\nu\alpha\beta}$, while $\phi(r)$ is the radial-dependent scalar field and $\lambda$ is the Gauss-Bonnet coupling constant with dimensions of \textit{length}.

The scalarization of electrically charged black holes is explored in \cite{Doneva:2018rou}. There exists a lower bound on the dimensionless Gauss-Bonnet coupling constant $\Tilde{\lambda}=\lambda/M$, which ensures the tachyonic instability of the Reissner-Nordstr\"{o}m (RN) black hole solution as the vacuum solutions of the theory.

We consider a static and spherically symmetric ansatz for the metric as follows:
\begin{equation}
    \mathrm{d}s^2=-e^{2\Phi(r)}\mathrm{d}t^2+e^{2\Lambda(r)}\mathrm{d}r^2+r^2\mathrm{d}\Omega^2.
\end{equation}
The Maxwell equation takes the form 
$V''(r)+V'(r)\left(\frac{2}{r}-\Lambda'(r)-\Phi'(r)\right)=0$.
Solving this equation shows that the electric charge $Q$ appears as an integration constant in the first derivative of the electric potential,
$V'(r)=e^{\Lambda+\Phi}Q/r^2$ \footnote{In \cite{Doneva:2018rou}, the construction of charged scalarized black hole solutions assumes a Coulombic form for the electromagnetic potential, $V(r)=-Q/r$. This expression solves the Maxwell equation only when the metric functions satisfy $\Lambda=-\Phi$, which constitutes a good approximation in the weak-coupling regime. In the present work, we go beyond this approximation by solving simultaneously for the spacetime metric, the scalar field, and the electromagnetic potential. This procedure ensures the full consistency of the coupled field equations and allows us also to explore the strong-coupling regime.}
The Einstein field equations of the theory are:
\begin{align}
    &\frac{2}{r}\left(1+\frac{2}{r}\left(1-3e^{-2\Lambda}\right)\Psi_r\right)\frac{d\Lambda}{dr}+\frac{e^{2\Lambda}-1}{r^2}-\frac{4}{r^2}\left(1-e^{-2\Lambda}\right)\frac{d\Psi_r}{dr}-\left(\frac{d\phi}{dr}\right)^2-e^{2\Lambda}\frac{Q^2}{r^4}=0,\label{feq1}\\
    &\frac{2}{r}\left(1+\frac{2}{r}\left(1-3e^{-2\Lambda}\right)\Psi_r\right)\frac{d\Phi}{dr}-\frac{e^{2\Lambda}-1}{r^2}-\left(\frac{d\phi}{dr}\right)^2+e^{2\Lambda}\frac{Q^2}{r^2}=0,\label{feq2}\\
    &\frac{d^2\Phi}{dr^2}+\left(\frac{d\Phi}{dr}+\frac{1}{r}\right)\left(\frac{d\Phi}{dr}-\frac{d\Lambda}{dr}\right)+\frac{4e^2\Lambda}{r}\left(3\frac{d\Phi}{dr}\frac{d\Lambda}{dr}-\frac{d^2\Phi}{dr^2}-\left(\frac{d\Phi}{dr}\right)^2\right)\Psi_r-\frac{4e^{-2\Lambda}}{r}\frac{d\Phi}{dr}\frac{d\Psi_r}{dr}+\left(\frac{d\phi}{dr}\right)^2-e^{2\Lambda}\frac{Q^2}{r^4}=0,\label{feq3}\\
    &\frac{d^2\phi}{dr^2}+\left(\frac{d\Phi}{dr}-\frac{d\Lambda}{dr}+\frac{2}{r}\right)\frac{d\phi}{dr}-\frac{2\lambda^2f'(\phi)}{r^2}\left(\left(1-e^{-2\Lambda}\right)\left(\frac{d^2\Phi}{dr^2}+\frac{d\Phi}{dr}\left(\frac{d\Phi}{dr}-\frac{d\Lambda}{dr}\right)\right)+3e^{-2\Lambda}\frac{d\Phi}{dr}\frac{d\Lambda}{dr}\right)=0,\label{eom}
\end{align}
where 
\begin{equation}
    \Psi_r=\lambda^2 f'(\phi)\frac{d\phi}{dr}.
\end{equation}
In our analysis, we adopt the Gaussian coupling model, which has been shown to yield dynamically stable and well-behaved solutions \cite{BlazquezSalcedo:2022omw}. The corresponding coupling function is given by:
\begin{equation}
    f(\phi)=\frac{1}{2\beta} \left(1-e^{-\beta\phi^2}\right),\label{gaussiancoupling}
\end{equation}
where $\beta$ is a dimensionless constant that controls the strength of the coupling between the scalar field and the Gauss-Bonnet invariant. Note that since $\beta$ appears in the denominator, small values of $\beta$ correspond to a strong effective coupling, while large values of $\beta$ correspond to a weak coupling. The conditions $f'(\phi)|_{\phi\to 0}=0$ and $f''(\phi)|_{\phi\to 0}>0$ are satisfied, ensuring, respectively, that the trivial black hole solutions with $\phi=0$ solve the field equations, and that tachyonic instabilities are present. The latter signals the existence of nontrivial scalarized black hole solutions.

The scalarized black hole solutions are constructed using a shooting method, with boundary conditions imposed by the requirements of asymptotic flatness and regularity at the black hole horizon $r=r_H$. 

The asymptotic solutions near the black hole horizon $r=r_H$ are derived under the assumption of the existence of a black hole horizon, meaning that the metric functions behave as $e^{2\Phi}|_{r\to r_H}\to 0$, $e^{-2\Lambda}|_{r\to r_H}\to 0$:
\begin{align}
    e^{2\Phi}\Big|_{r\to r_H}=& a_1 (r-r_H)+a_2(r-r_H)^2+\dots,\nonumber\\
    e^{2\Lambda}\Big|_{r\to r_H}=&b_0+\frac{b_1}{(r-r_H)}+\frac{b2}{(r-r_H)^2}+\dots,\nonumber\\
    \phi\Big|_{r\to r_H}=& \phi_H+\phi'_H (r-r_H)+\phi''_H(r-r_H)^2+\dots. \label{asymptoticbehaviorathorizon}
\end{align}
These asymptotics yield the conditions for the first derivatives of the functions as:
\begin{align}
    \Phi'\Big|_{r\to r_H}=&\frac{1}{2}\frac{1}{(r-r_H)}+\dots,\nonumber\\
    \Lambda'\Big|_{r\to r_H}=&-\frac{1}{2}\frac{1}{(r-r_H)}+\dots,\nonumber\\
    \phi'\Bigg|_{r\to r_H}=&\frac{1}{4 \lambda ^2 r_H f'(\phi) \left(-4 \lambda ^4 Q^2 f'(\phi_H)^2-Q^2 r_H^4+r_H^6\right)}\left(4 \lambda ^4 Q^2 \left(Q^2+r_H^2\right) f'(\phi_H)^2+\left(Q^2- r_H^2\right)\left(r_H^6\right.\right.\nonumber\\
    &\left.\left.\pm\sqrt{16 \lambda ^8 Q^2 \left(Q^2+6 r_H^2\right) f'(\phi_H)^4+8 \lambda ^4 r_H^6 \left(2 Q^2-3 r_H^2\right) f'(\phi_H)^2+r_H^{12}}\right)\right)\label{boundaryscalar}
\end{align}
We focus on the plus sign, as this choice recovers the RN black hole solution in the absence of a scalar field, as long as it recovers the respective condition for the neutral case ($Q=0$), \cite{Doneva:2017bvd}. This expression indicates that non-trivial and regular scalar field configurations can exist only if the quantity under the square root is positively defined in (\ref{boundaryscalar}):
\begin{equation}
    16 \lambda ^8 Q^2 \left(Q^2+6 r_H^2\right) f'(\phi_H)^4+8 \lambda ^4 r_H^6 \left(2 Q^2-3 r_H^2\right) f'(\phi_H)^2+r_H^{12}>0. \label{conditionofscalarization}
\end{equation}

The asymptotic behavior of solutions at large radial distances is obtained by imposing asymptotic flatness, that is, the scalarized spacetime approaches Minkowski spacetime as $r\to\infty$. Expanding all fields in inverse powers of the radial coordinate, $1/r$, one finds
\begin{align}
    \Lambda\Big|_{r\to\infty}=& \frac{M}{r}+\frac{2M^2-Q^2-D^2}{2r^2}
    +\frac{8M^3-6MQ^2-9MD^2}{6r^3}+\dots ,\nonumber\\
    \Phi\Big|_{r\to\infty}=& -\frac{M}{r}-\frac{2M^2-Q^2}{2r^2}
    -\frac{8M^3-6MQ^2-MD^2}{6r^3}+\dots ,\nonumber\\
    \phi\Big|_{r\to\infty}=& \frac{D}{r}+\frac{DM}{r^2}
    +\frac{8DM^2-2DQ^2-D^3}{6r^3}+\dots ,\nonumber\\
    V\Big|_{r\to\infty}=& \Phi_e-\frac{Q}{r}+\dots ,
    \label{asymptoticbehavioratinfty}
\end{align}
where $M$, $Q$, $\Phi_e$, and $D$ denote, respectively, the Arnowitt-Deser-Misner (ADM) mass, the electric charge, the electric potential at infinity, and the scalar charge of the scalarized black hole solution. The scalar charge $D$ is a secondary hair, since it is not associated with an independent conserved quantity but is uniquely determined by the boundary conditions \cite{Ballesteros:2025wvs}. 

The scalarized solutions exhibit entropy that deviates from the Area Law, following Wald's formula \cite{Wald:1993nt}:
\begin{equation}
    S_H=\frac{A_H}{4}+4\pi\lambda^2 f(\phi_H). \label{entropyWald}
\end{equation}
The temperature of spherical symmetric scalarized black holes is derived from the surface gravity $k_H$ at the horizon.
\begin{equation}
    T_H=\frac{k_H}{2\pi}=\frac{1}{4\pi}\left(\frac{1}{\sqrt{|g_{tt}g_{rr}|}}\left|\frac{dg_{tt}}{dr}\right|\right)_{r_H}=\dfrac{1}{4\pi}\sqrt{\dfrac{a_1}{b_1}}.\label{temperature}
\end{equation}

In theories that admit scalarized solutions, the comparison between the entropy of general relativistic solutions and that of the corresponding scalarized black holes is often used to assess the thermodynamic preference of the latter, as they typically possess a larger entropy \cite{Doneva:2022ewd}. This comparison refers to the microcanonical ensemble in which the mass $M$ and the electric charge $Q$ are held fixed. However, thermodynamic stability is ensemble-dependent: a configuration that is stable in the microcanonical ensemble need not be stable in other ensembles, such as the canonical or grand canonical ones. This is because different ensembles impose different constraints on the allowed fluctuations, which in turn modify the relevant thermodynamic potentials and stability criteria \cite{Wald:1999vt}.

\section{Euclidean thermodynamics and thermal states} \label{euclidean thermodynamics}
In this section, we investigate the thermodynamics of scalarized black holes in the EMsGB theory using the Euclidean approach in the canonical ensemble, where the temperature $T$ and the electric charge $Q$ are held fixed. The Euclidean continuation of the static and spherically symmetric black hole ansatz reads:
\begin{equation}
\mathrm{d} s^2 = N(r)^2 e^{2\Phi(r)}\, \mathrm{d}\tau^2 + e^{2\Lambda(r)}\, \mathrm{d}r^2 + r^2 \mathrm{d}\Omega^2 ,
\end{equation}
where $N(r)$ denotes the lapse function and $\tau = i t$ is the Euclidean time. The Euclidean time coordinate is periodically identified as $0 \leq \tau \leq \beta_\tau$, with the period $\beta_\tau$ fixed by the requirement of regularity of the Euclidean section at the black hole horizon, namely by the absence of conical singularities. This condition leads to
\begin{equation}
\beta_\tau = 4\pi \left.\frac{\sqrt{g_{\tau\tau}(r)\, g_{rr}(r)}}{\partial_r g_{\tau\tau}(r)}\right|_{r=r_H}.
\end{equation}

In gravitational thermodynamics, the temperature $T$ is fixed by the periodicity condition of the Euclidean time coordinate $T=1/\beta_\tau$. 
The Euclidean action, introduced in Eq.~(\ref{action}), can be written as:
\begin{align}
  \mathcal{I}_E=\beta_\tau \int_{r_H}^\infty \mathrm{d}r \left(NH-A_\tau \partial_r\pi^r\right) +\mathcal{B},\label{euclidean action}
\end{align}
where $\mathcal{B}$ is a boundary term that ensures a well-defined variational problem and 
\begin{align}
    H=&-\frac{e^{-\Lambda-\Phi} (\pi^r)^2}{2r^2}+e^{-\Lambda+\Phi}\left(\frac{1-e^{2\Lambda}-r\Lambda'+r^2\phi'}{2}-2\Psi_r \Lambda'(1-3e^{-2\Lambda})+2\frac{d\Psi_r}{dr}(1-e^{-2\Lambda})\right). \label{hamiltonian constraint}
\end{align}
The $\pi^r$ stands for the non-vanishing component of the electromagnetic conjugate momentum defined as:
\begin{equation}
    \pi^r=\frac{e^{-\Lambda-\Phi} r^2 V'}{N}
\end{equation}
The variation of the Euclidean action with respect to all the dynamical fields yields:
\begin{align}
   \delta \mathcal{I}_E=&\beta_\tau \int\,\mathrm{d}r\, \left(E_N\delta N+E_\Phi\delta\Phi +E_\Lambda\delta\Lambda+E_\phi\delta\phi+E_V\delta V+E_{\pi^r}\delta\pi^r\right) \nonumber\\
   &+\left(Ne^{-3\Lambda+\Phi}\delta\Lambda(2\Psi_r(3-e^{2\Lambda})-re^{2\Lambda})\right)\Big|_{r_H}^\infty\nonumber\\
   &+\left(e^{-3\Lambda+\Phi} \delta\phi \left(e^{2\Lambda}r^2N\phi'+2\lambda^2(1-e^{2\Lambda})(f'N'+Nf'\Phi'-N\phi'f'')\right)\right)\Big|_{r_H}^\infty\nonumber\\
   &-\left(2e^{-3\Lambda+\Phi}\lambda^2Nf'(1-e^{2\Lambda}\delta\phi')\right)\Big|_{r_H}^\infty-\left(V\delta\pi^r\right)\Big|_{r_H}^\infty +\delta\mathcal{B},
\end{align}
where $E_i$ represent the Euler-Lagrange equations as follows:
\begin{align}
    E_N      &:\left(1+2(1-3e^{-2\Lambda})\Psi_r\right)\Lambda'+\frac{e^{2\Lambda}-1}{2}+2(1-e^{-2\Lambda})\Psi_r'+\frac{r^2\phi'^2}{2}+\frac{e^{-2\Phi}(\pi^r)^2}{2r^2}=0  \label{EN}   \\
    E_\Phi   &:-\left(r+2(1-3e^{-2\Lambda})\Psi_r'\right)\Lambda'+\frac{1-e^{2\Lambda}}{2}+\frac{r^2{\phi'}^2}{2}+2(1-e^{-2\Lambda})\Psi_r'=0 \label{EPhi}\\
    E_\Lambda &:(N\Phi'+N')\left(r+2(1-3e^{-2\Lambda})\Psi_r\right)+\frac{N(1-e^{2\Lambda})}{2}-\frac{r^2N{\phi'}^2}{2} +\frac{e^{-2\Phi} N{\pi^r}^2}{2r^2} =0 \label{ELambda}\\
    E_\phi   & :N\left(-r^2\phi''-r(r(\Phi'-\Lambda')+2)\phi'+2\lambda^2f'\left((1-3^{-2\Lambda})(\Phi''+\Phi'(\Phi'-\Lambda'))+2e^{-2\Lambda}\Lambda'\Phi'\right)\right)\nonumber\\
    &\phantom{aa}N'\left(2\lambda^2f'\left(2\Phi'(1-e^{-2\Lambda})-\Lambda'(1-3e^{-2\Lambda})\right)-r^2\phi'\right)+2\lambda^2f'N''(1-e^{-2\Lambda})=0\label{Ephi}\\
    E_V      & : {\pi^r}'=0\label{EV}\\
    E_{\pi^r}& : V'(r)-\frac{N\pi^r e^{-\Lambda-\Phi}}{r^2}=0 . \label{Epir}
\end{align}
The lapse function $N$ appears as a Lagrange multiplier in the Euclidean action Eq.~(\ref{euclidean action}). Varying the action with respect to $N$ yields Eq.~(\ref{EN}), which can be identified as the Hamiltonian constraint, Eq.~(\ref{hamiltonian constraint}). Consequently, without loss of generality, the gauge may be fixed by setting the lapse function to unity $N=1$.

Eq.~(\ref{EV}) corresponds to the Gauss law, from which it follows that the radial component of the conjugate momentum, $\pi^r$, is constant. Due to the Wick rotation, the conjugate momentum is taken to be imaginary, $\pi^r=i Q$, where $Q$ is an integration constant. Substituting this expression into the remaining Eq.~(\ref{EN}), (\ref{EPhi}), (\ref{ELambda}), and (\ref{Ephi}), yields the corresponding system of field equations in the Lorentzian sector, namely Eq.~(\ref{feq1}), (\ref{feq2}), (\ref{feq3}), and (\ref{eom}).

Therefore, on shell, the scalarized black hole solutions derived in Sec.~\ref{thescalarization} satisfy the above-mentioned field equations $E_i$, and consequently contribute to the variation of the boundary term, denoted by $\delta\mathcal{B}$, as
\begin{equation}
    \delta\mathcal{B}=\beta_\tau\left(\delta M-T\left(\frac{\delta A}{4}+4\pi\lambda^2\delta f(\phi_H)\right)-\Phi_e \delta Q\right).\label{boundaryterm}
\end{equation}

In deriving this expression, we have used the asymptotic behavior of the functions near the horizon and at infinity, Eqs.~(\ref{asymptoticbehaviorathorizon}) and (\ref{asymptoticbehavioratinfty}), together with the expression for the temperature, Eq.~(\ref{temperature}), and the variation of the horizon area $\delta A = 8\pi r_H \delta r_H$.

The thermodynamic potential $G$ can then be obtained through the quantum statistical relation:
\begin{equation}
    \mathcal{I}_E=\frac{G}{T}\Rightarrow 
    G= M-T\left(\frac{ A}{4}+4\pi \lambda^2  f(\phi_H)\right)-\Phi_e  Q .
\end{equation}
Here, the entropy $S$ does not follow the standard area law, in agreement with the result obtained from the Wald entropy formula, described by Eq.~ (\ref{entropyWald}).

Thus, the on-shell action with a well-posed variational principle defines the grand canonical ensemble, in which the electric potential $\Phi_e$ is held fixed, and is associated with the Gibbs free energy:
\begin{equation}
G = M - TS - \Phi_e Q \,.
\end{equation}

Fixing the electric charge $Q$ at the boundary instead requires a Legendre transformation of the action, leading to the canonical ensemble, which is described by the Helmholtz free energy:
\begin{equation}
F = M - TS \,.
\label{freeenergy}
\end{equation}

Note that, since our solutions are obtained numerically, the first law of black hole thermodynamics can be inferred by imposing $\delta \mathcal{B}=0$. From Eq.~(\ref{boundaryterm}), this condition leads to:
\begin{equation}
\delta M = T \delta S + \Phi_e \delta Q .
\end{equation}
The secondary scalar charge $D$ does not introduce an additional thermodynamic variable in the first law. Nevertheless, it leaves a clear imprint on the thermodynamic properties of the scalarized black hole branches.

In the following, we focus on thermal states with fixed electric charge $Q$ and adopt dimensionless thermodynamic quantities, defined as:
\begin{equation}
    \Phi_e\to \Phi_e, \phantom{aa} Q\to Q/\lambda, \phantom{aa} T_H\to T_H\lambda, \phantom{aa} S_H\to S_H/\lambda^2, \phantom{aa} F\to F/\lambda. 
\end{equation}

\section{Thermodynamical analysis}\label{Thermodynamical analysis}

The quantity that characterizes the local thermodynamic stability or instability of a thermal state is the specific heat, defined as $C_Q=\partial M/\partial T$. When the specific heat of a black hole is negative, $C<0$, small thermal fluctuations drive the system away from equilibrium: the black hole may absorb energy and decrease its temperature, causing further heat inflow and runaway growth, or it may release energy and increase its temperature, leading to heat outflow and runaway evaporation. Consequently, such configurations cannot remain in stable thermal equilibrium. By contrast, a positive specific heat, $C>0$, indicates local thermodynamic stability.

In the present work, since our black hole solutions are obtained numerically, we do not compute the specific heat directly. Instead, we make use of the first law of black hole thermodynamics, which implies $C_Q=T(\partial S/\partial T)$. Since the Hawking temperature is always positive, the sign of the specific heat is completely determined by the slope of the equation of state $T(S)$: an increasing function $T(S)$ corresponds to a positive specific heat, whereas a decreasing function corresponds to a negative one \cite{Chatzifotis:2023ioc}.

In the canonical ensemble, global thermodynamic stability is determined by minimizing the Helmholtz free energy at fixed temperature and charge, so that the physically realized configuration corresponds to the state with the lowest value of $F$ \cite{Hawking:1982dh}. Phase transitions between two thermal states, denoted by $A$ and $B$, can be identified from the free energy diagram $F$ as a function of temperature $T$. A transition occurs at a critical temperature $T_{\mathrm{PT}}$, where the two branches exchange global dominance. The order of the transition is determined by the behavior of the free energy and its derivatives with respect to temperature. At $T_{\mathrm{PT}}$, if $F_A \neq F_B$, the transition is referred to as a zeroth-order phase transition, characterized by a finite jump in the free energy itself. If $F_A = F_B$ but $\partial F_A/\partial T \neq \partial F_B/\partial T$, the thermodynamic relation $S = -\partial F/\partial T$ implies $S_A \neq S_B$, corresponding to a first-order phase transition. Finally, if $F_A = F_B$, $S_A = S_B$, but $\partial^2 F_A/\partial T^2 \neq \partial^2 F_B/\partial T^2$, i.e.\ $C_{Q,A} \neq C_{Q,B}$ (since $C_Q = -T\,\partial^2 F/\partial T^2$), the transition is of second order \cite{Landau:1980mil}.

According to the above considerations, we explore the local and global thermodynamic stability of the thermal states described by the RN and charged scalarized black hole solutions. In particular, we analyze the phase structure for a sequence of values of the coupling constant $\beta$, covering both the weak- and the strong-coupling regimes. For illustrative purposes, we present results for the representative case $Q/\lambda = 0.4$; any other value of this ratio yields qualitatively equivalent behavior.

\begin{figure}[h]  
\centering
    \includegraphics[width=0.45\textwidth]{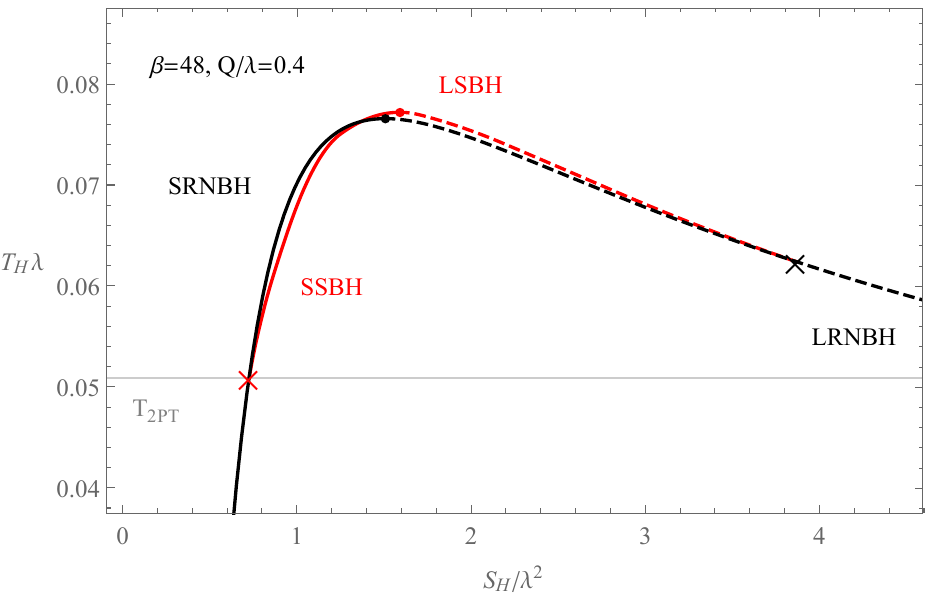} 
    \includegraphics[width=0.44\textwidth]{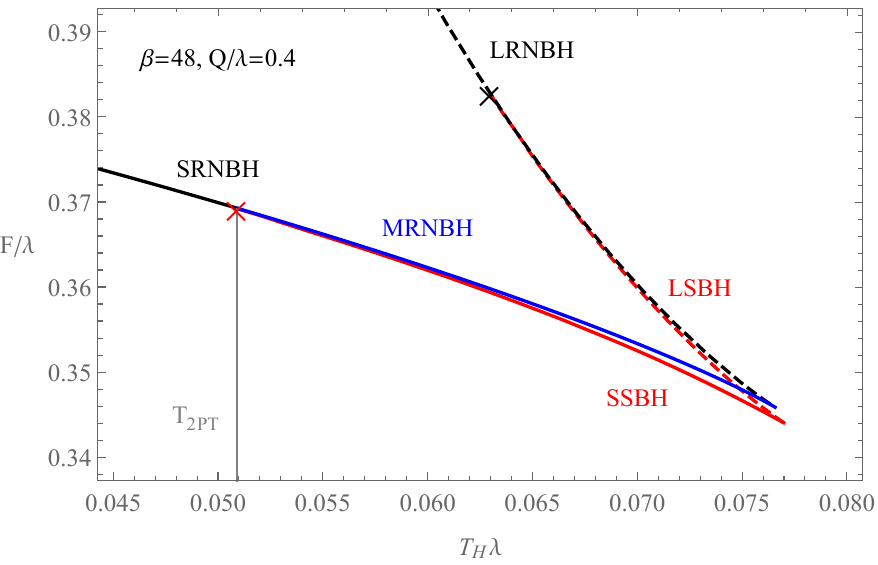}
    \caption{Thermodynamic behavior of the solutions for $\beta=48$. Left panel: Equation of state given by the temperature $T_H$ as a function of the entropy $S_H$, showing isocharged curves. Right panel: Helmholtz free energy $F$ as a function of the temperature $T_H$. In both panels, the dashed black curves correspond to the large, locally unstable RN branch (LRNBH), while the solid black curves represent the stable small RN black holes (SRNBH). The dashed red curves denote the large, locally unstable scalarized black holes (LSBH), whereas the solid red curves correspond to the stable small scalarized configurations (SSBH). The blue curves indicate metastable RN phases (MRNBH). The gray solid line marks the second-order phase transition. Crosses indicate bifurcation points, while dots mark the Davies points.}
    \label{thermanalysisbeta48}
\end{figure}

In Fig.~\ref{thermanalysisbeta48}, we present the case $\beta = 48$, which corresponds to the weak-coupling regime. The left panel shows the equation of state $T(S)$ along isocharged curves for each family of solutions. The RN branch (shown by the black curve) exhibits a Davies point (indicated by the black dot) where the temperature reaches a maximum, and the specific heat diverges. Consequently, small RN black holes (represented by the black solid curve, SRNBH), possess a positive specific heat and are locally thermodynamically stable, whereas large RN black holes (represented by the black dashed curve, LRNBH), have negative specific heat and are therefore locally unstable. The charged scalarized branch (shown by the red curve), bifurcates from the trivial RN branch at higher masses/entropies (indicated by the black cross), and extends toward lower masses/entropies until it reconnects with the RN branch at a second bifurcation point (marked by the red cross). This second bifurcation corresponds dynamically to a spontaneous loss of scalar hair before the extremal limit $T=0$ is reached, a feature already reported in \cite{Doneva:2018rou}. The scalarized branch also exhibits a Davies point (indicated by the red dot), where the specific heat diverges. Accordingly, small scalarized black holes (represented by the red solid curve, SSBH), are locally stable, while large scalarized black holes (shown by the red dashed curve, LSBH), are locally unstable.

A snapshot of the Helmholtz free energy is shown in the right panel of Fig.~\ref{thermanalysisbeta48}. Both the RN and scalarized branches exhibit Davies points, which appear in the free-energy diagram as cusps and signal the maximum temperature at which each family of thermal configurations can exist. 

There exists a second-order phase transition between the SRNBH and SSBH branches. As can be inferred from both panels of Fig.~\ref{thermanalysisbeta48}, at the transition temperature $T_{2\mathrm{PT}}$ (indicated by the horizontal and vertical solid lines in the left and right panel, respectively) the free energies and entropies of the two locally stable branches coincide, $F_{2\mathrm{PT}}^{\mathrm{SRNBH}} = F_{2\mathrm{PT}}^{\mathrm{SSBH}}$ and $S_{2\mathrm{PT}}^{\mathrm{SRNBH}} = S_{2\mathrm{PT}}^{\mathrm{SSBH}}$, while their heat capacities differ, $C_{Q,2\mathrm{PT}}^{\mathrm{SRNBH}} \neq C_{Q,2\mathrm{PT}}^{\mathrm{SSBH}}$. This means that an SSBH created in this region will cool as it evaporates, undergoing a second-order phase transition to the trivial SRNBH configuration at $T_{2\mathrm{PT}}$. The behavior indicates that the second bifurcation point that signals the offset of scalarization is associated with a second-order phase transition between the two thermodynamically stable branches. For temperatures above the transition temperature, $T > T_{2\mathrm{PT}}$, a segment of the locally stable small RN black hole branch corresponds to metastable thermal states (denoted by the blue curve, MRNBH), as these configurations do not minimize the free energy globally.

\begin{figure}[h]  
\centering
    \includegraphics[width=0.45\textwidth]{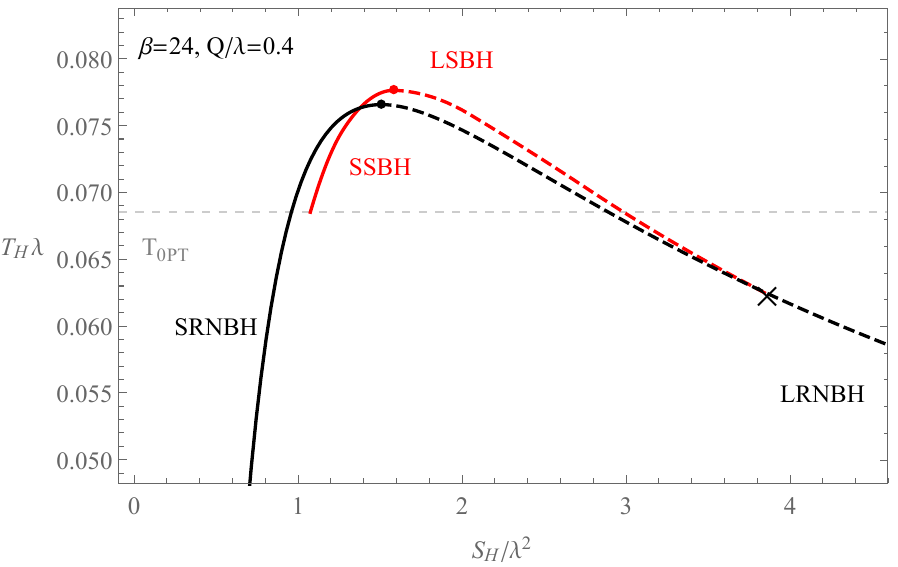} 
    \includegraphics[width=0.44\textwidth]{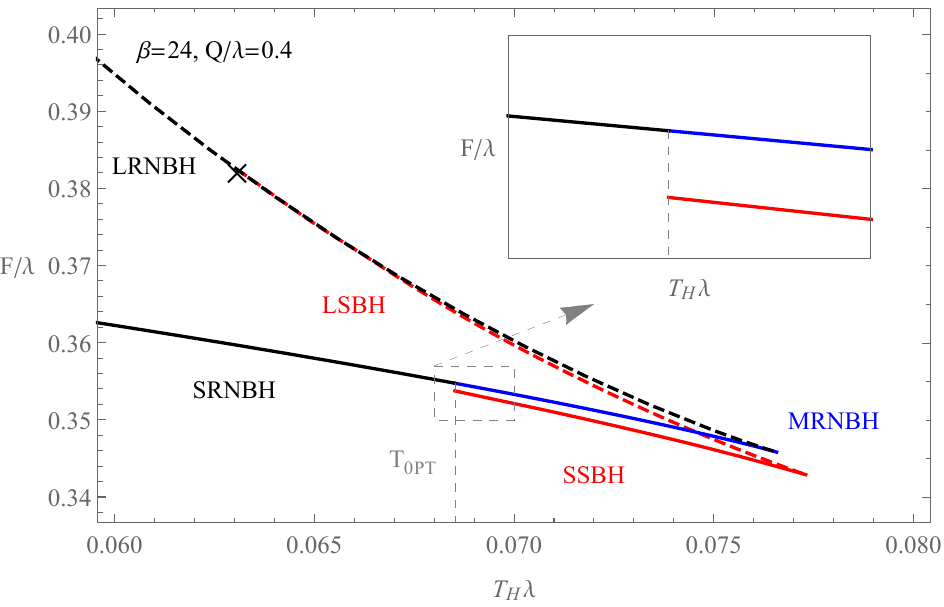}
    \caption{Thermodynamic behavior of the solutions for $\beta=24$. The curves and symbols follow the same conventions as in Fig.~\ref{thermanalysisbeta48}. The dashed lines indicate the zeroth-order phase transition.}
    \label{thermanalysisbeta24}
\end{figure}

The behavior of the thermal states for a higher value of the coupling, $\beta = 24$, is shown in Fig.~\ref{thermanalysisbeta24}. In this regime, the charged scalarized branch shrinks,as the field equations cease to support nontrivial solutions with the required boundary conditions, and terminates before reaching a second bifurcation point with the RN branch, so that the two stable phases, SRNBH and SSBH, are no longer connected by a continuous second-order transition as in the previous case. Instead, spontaneous loss of scalar hair is accompanied by a finite jump in the free energy, $F_{0\mathrm{PT}}^{\mathrm{SRNBH}} \neq F_{0\mathrm{PT}}^{\mathrm{SSBH}}$, and the transition between the two stable branches becomes of the zeroth order, occurring at a temperature $T_{0\mathrm{PT}}$ (indicated by the horizontal and vertical dashed lines in the left and right panel, respectively). An SSBH will cool as it evaporates and eventually undergo a zeroth-order phase transition to an SRNBH. For temperatures $T > T_{0\mathrm{PT}}$, the locally stable RN black holes correspond to metastable thermal states.

\begin{figure}[h]  
\centering
    \includegraphics[width=0.45\textwidth]{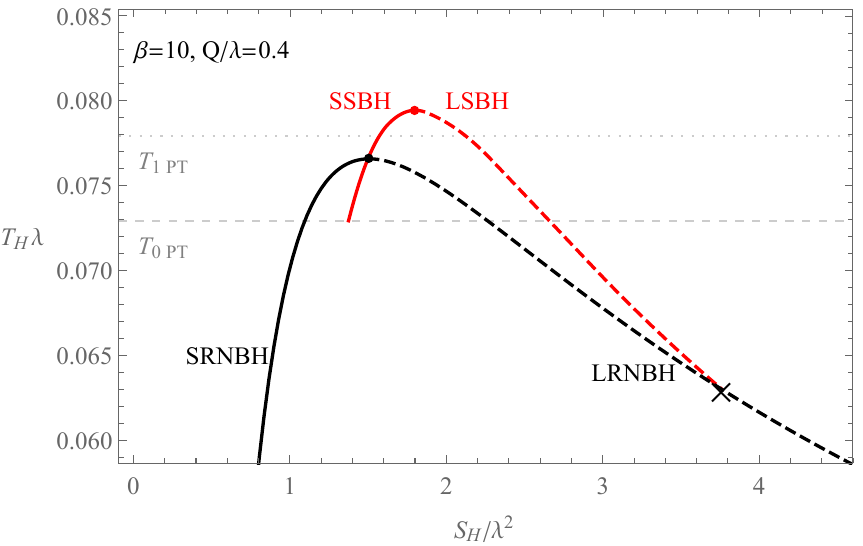} 
    \includegraphics[width=0.44\textwidth]{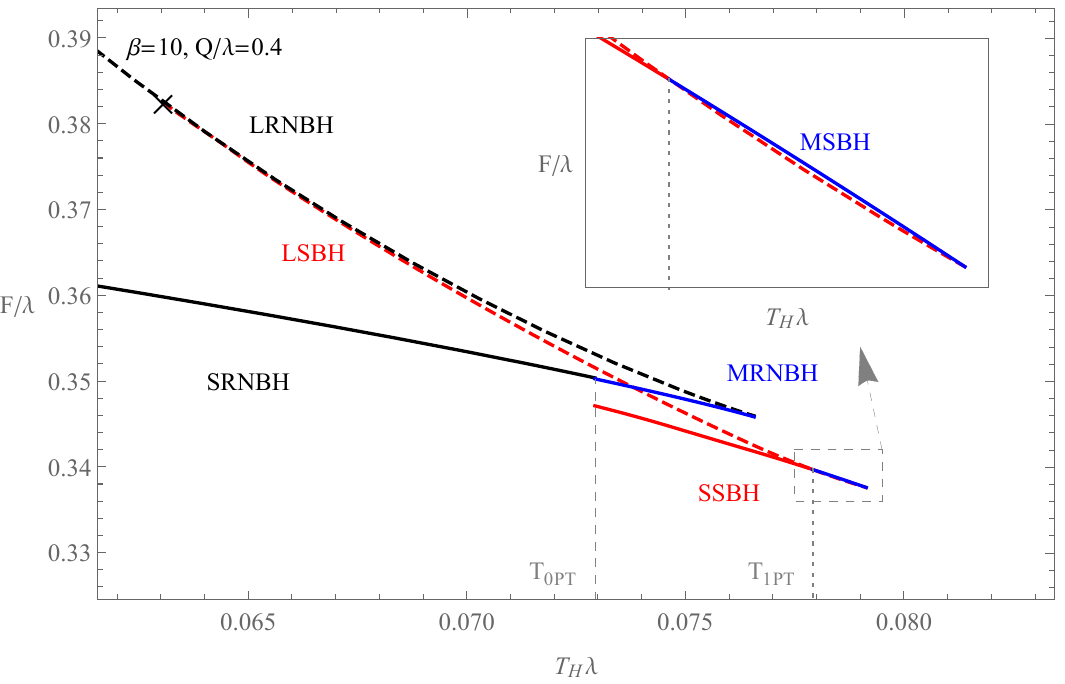}
    \caption{Thermodynamic behavior of the solutions for $\beta=10$. The curves and symbols follow the same conventions as in Fig.~\ref{thermanalysisbeta24}. The dotted line indicates the first-order phase transition.}
    \label{thermanalysisbeta10}
\end{figure}

In Fig.~\ref{thermanalysisbeta10}, we present the case $\beta = 10$, where a more pronounced modification of the phase structure emerges. In the left panel, we show the equation of state $T(S)$, where the scalarized branch reaches higher temperatures than in the previous cases while simultaneously shrinking in extent. The right panel presents the corresponding free-energy diagram. In contrast to the previous case, the coupling is now sufficiently large that, in the vicinity of the Davies point of the scalarized branch, the locally thermodynamically stable configurations develop a metastable sector (indicated by the blue curve, MSBH). This feature induces a characteristic fish-like structure in the free-energy diagram, as there exists a finite temperature interval, from $T_{0\mathrm{PT}}$ to $T_{1\mathrm{PT}}$, over which scalarized black holes are thermodynamically favored, defining the SSBH phase.

The emergence of this fish-like structure highlights the nontrivial influence of the coupling parameter $\beta$ in determining the shape of the free-energy diagram associated with black hole configurations. Consequently, three thermodynamically dominant phases are identified: two locally stable configurations corresponding to the SRNBH and SSBH branches, and a phase with large scalarized black holes (denoted by the red dashed curve, LSBH) that coexist with the MSBH solutions. These three phases are delimited by two phase transitions: a zeroth-order phase transition at temperature $T_{0\mathrm{PT}}$ and a first-order transition at temperature $T_{1\mathrm{PT}}$ (indicated by the dotted line) where $F_{1\mathrm{PT}}^{\mathrm{SSBH}} = F_{1\mathrm{PT}}^{\mathrm{LSBH}}$ and $S_{1\mathrm{PT}}^{\mathrm{SSBH}} \neq S_{1\mathrm{PT}}^{\mathrm{LSBH}}$. 

A scalarized black hole starting on the LSBH branch cools as it accretes mass and eventually undergoes a first-order phase transition to the SSBH branch at $T_{1\mathrm{PT}}$. Beyond this transition, no further thermodynamic phase changes occur unless an external mechanism halts accretion and induces evaporation of the SSBH configuration. We note that such a scenario requires both accretion and subsequent evaporation, and therefore does not correspond to a natural astrophysical pathway but rather to a theoretically constructed thermodynamic process. Conversely, a scalarized black hole initially in the stable SSBH region evolves analogously to the $\beta = 24$ case: as it evaporates, it undergoes a zeroth-order phase transition to the SRNBH branch at $T_{0\mathrm{PT}}$.

\begin{figure}[h]  
\centering
    \includegraphics[width=0.45\textwidth]{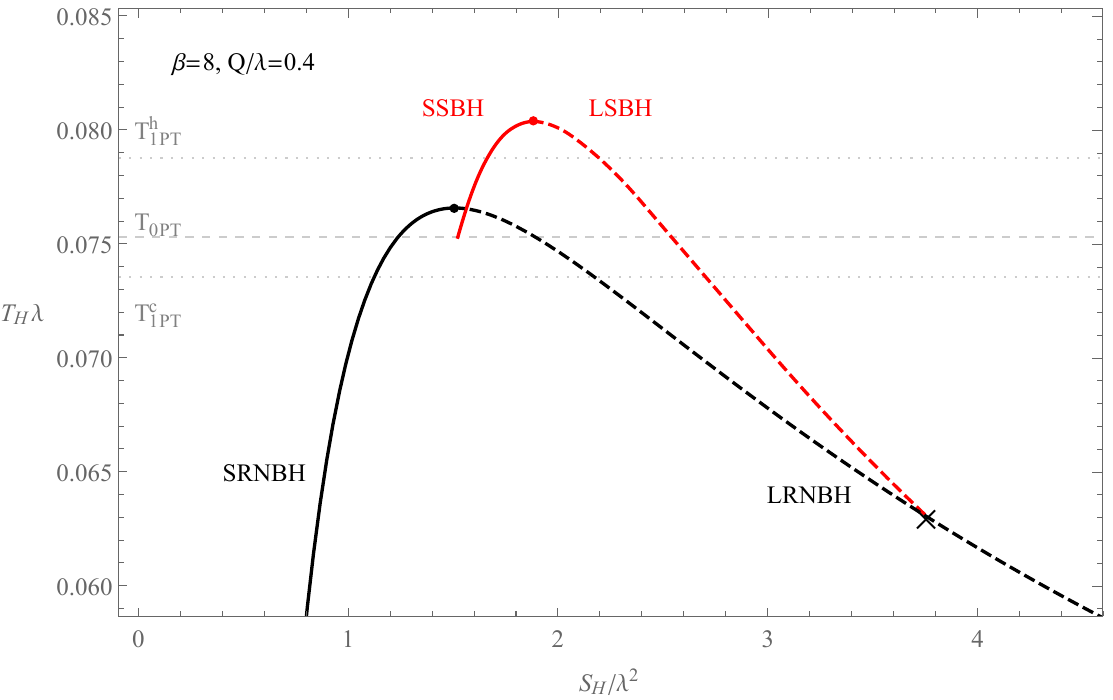} 
    \includegraphics[width=0.44\textwidth]{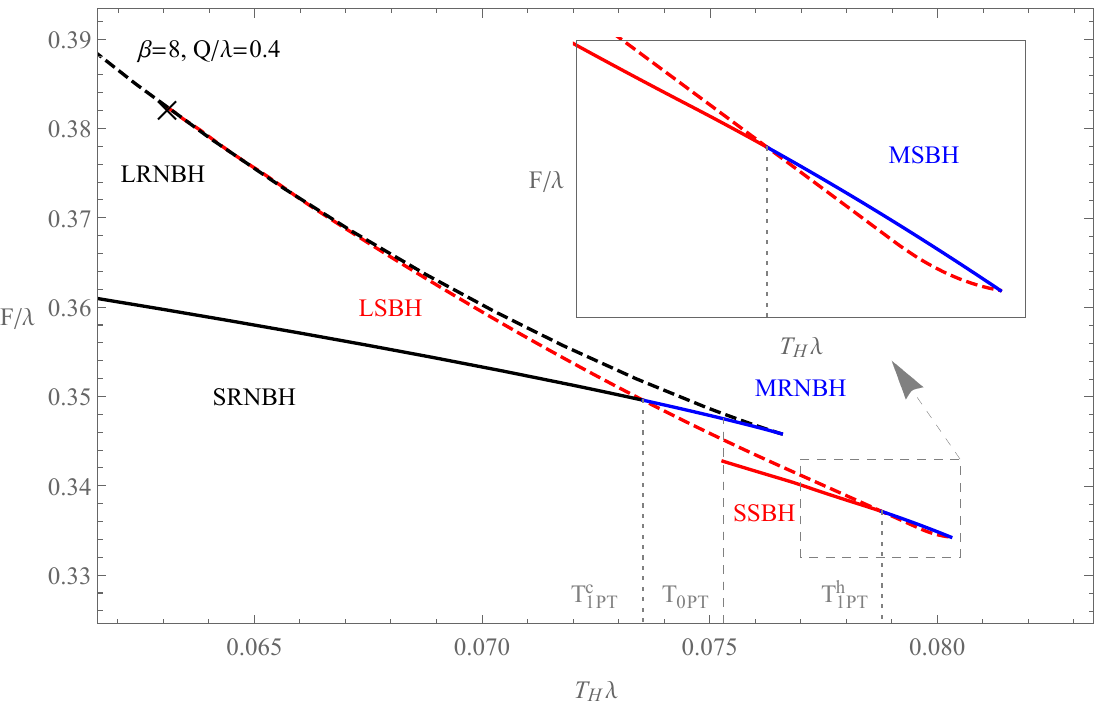}
    \caption{Thermodynamic behavior of the solutions for $\beta=8$. The curves and symbols follow the same conventions as in Fig.~\ref{thermanalysisbeta10}.}
    \label{thermanalysisbeta8}
\end{figure}

The case $\beta = 8$ is presented in Fig.~\ref{thermanalysisbeta8}. In this regime, the scalarized branch departs significantly from the RN branch and continues to shrink, as shown in the equation of state in the left panel. 

The corresponding free-energy diagram is displayed in the right panel. Compared to the previous case, the phase structure is now richer as it displays four thermodynamically preferred sectors: the system exhibits two locally stable regions associated with the SRNBH and SSBH branches, together with two locally unstable scalarized regions (indicated by the red dashed curves, LSBH). These correspond to a cold unstable region, defined by $T^c_{1\mathrm{PT}} < T < T_{0\mathrm{PT}}$, and a hot unstable region, defined by $T > T^h_{1\mathrm{PT}}$, where the last one coexists with metastable scalarized configurations (MSBH). The system therefore exhibits three phase transitions: two first-order transitions at temperatures $T^c_{1\mathrm{PT}}$ and $T^h_{1\mathrm{PT}}$, and one zeroth-order transition at the intermediate temperature $T_{0\mathrm{PT}}$, occurring between the cold LSBH and the SSBH branches. However, not all of these transitions can be achieved sequentially without external contrivance.

A scalarized black hole initially located in the cold LSBH region may evaporate, losing mass and consequently increasing its temperature. During this process, it moves rightward along the red dashed curve until it undergoes a zeroth-order phase transition to the SSBH branch at $T_{0\mathrm{PT}}$. After the transition, the resulting SSBH lies in a thermodynamically stable region and therefore represents the preferred equilibrium configuration. If instead the same unstable configuration accretes mass and grows, its temperature decreases as it moves left along the red dashed curve until it undergoes a first-order phase transition to the SRNBH branch at $T^c_{1\mathrm{PT}}$. Hence, locally unstable scalarized black holes in this region evolve toward a locally stable configuration regardless of whether they evaporate or accrete mass. The phase transition occurring in the hot LSBH region at $T^h_{1\mathrm{PT}}$ follows the same behavior described for the case $\beta = 10$: as it accretes mass, it undergoes a first-order phase transition to the SSBH branch at $T_{1\mathrm{PT}}$.

\begin{figure}[h]  
\centering
    \includegraphics[width=0.45\textwidth]{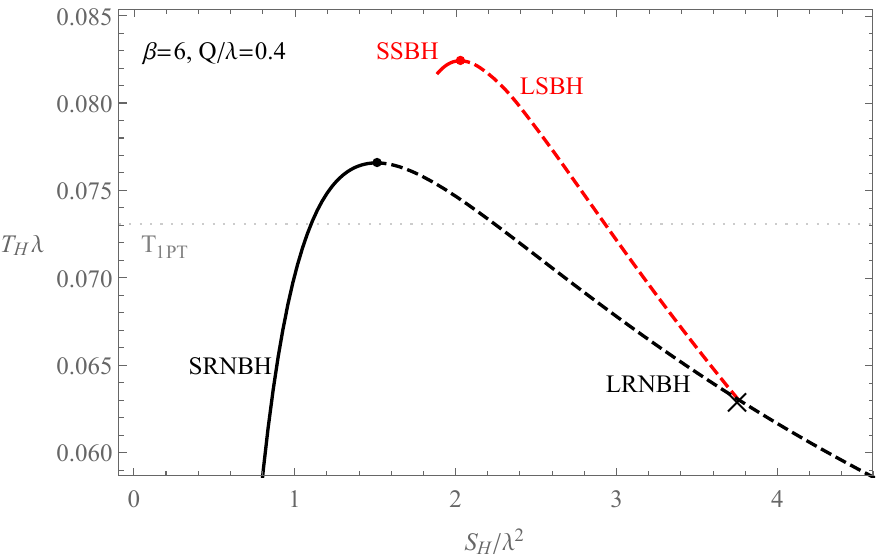} 
    \includegraphics[width=0.44\textwidth]{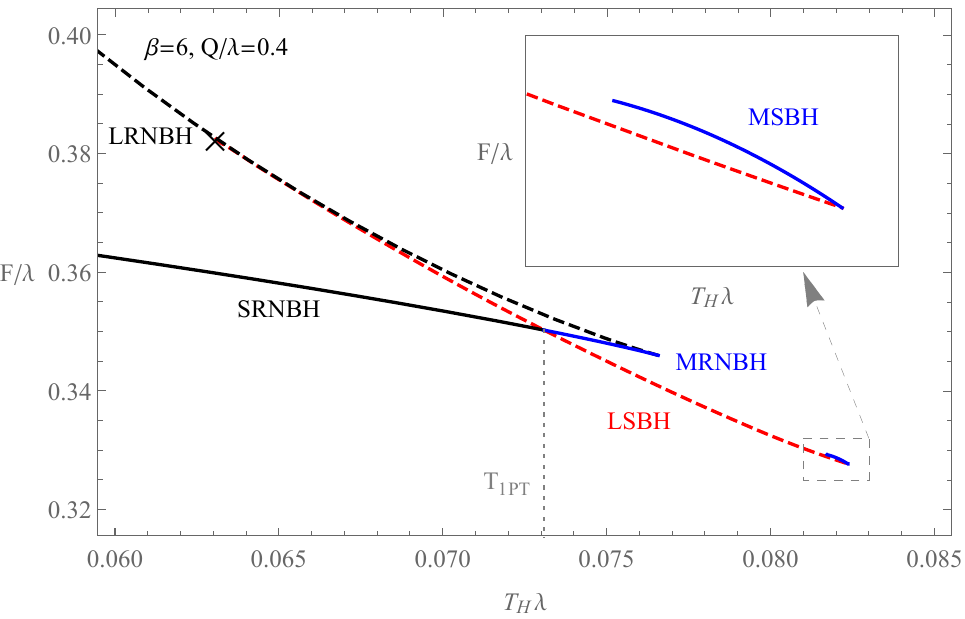}
    \caption{Thermodynamic behavior of the solutions for $\beta=6$. The curves and symbols follow the same conventions as in Fig.~\ref{thermanalysisbeta8}.}
    \label{thermanalysisbeta6}
\end{figure}

In Fig.~\ref{thermanalysisbeta6}, we present the coupling case $\beta = 6$, for which the locally stable scalarized branch becomes entirely metastable. In this regime, only a single first-order phase transition remains. This transition can be realized if a black hole configuration forms within the unstable region of the large scalarized branch and subsequently grows. As its mass increases, the temperature decreases, and the system evolves leftward along the red dashed curve until the first-order phase transition is reached at $T_{1\mathrm{PT}}$, where the configuration transitions to the SRNBH phase. The resulting larger SRNBH then ceases to grow as its positive specific heat allows it to reach thermal equilibrium.

\begin{figure}[h]  
\centering
    \includegraphics[width=0.45\textwidth]{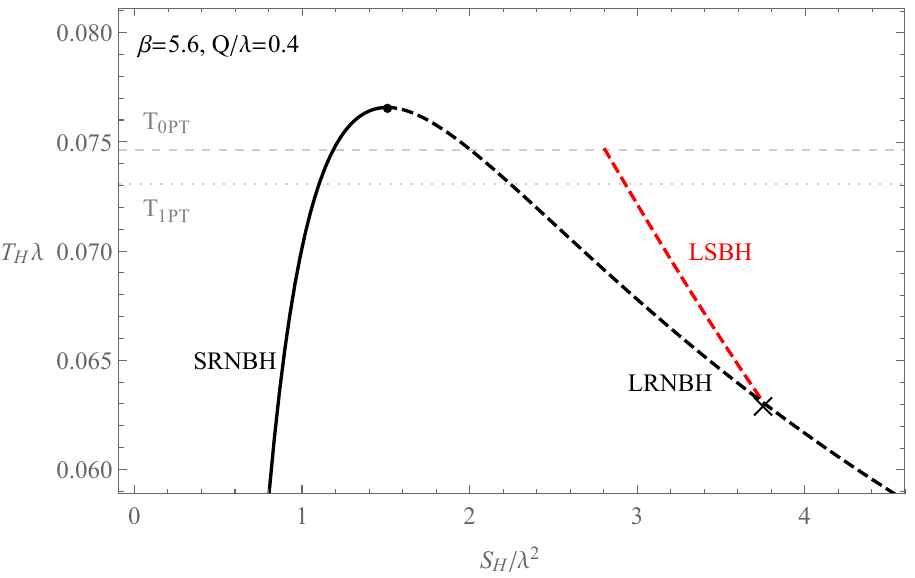} 
    \includegraphics[width=0.44\textwidth]{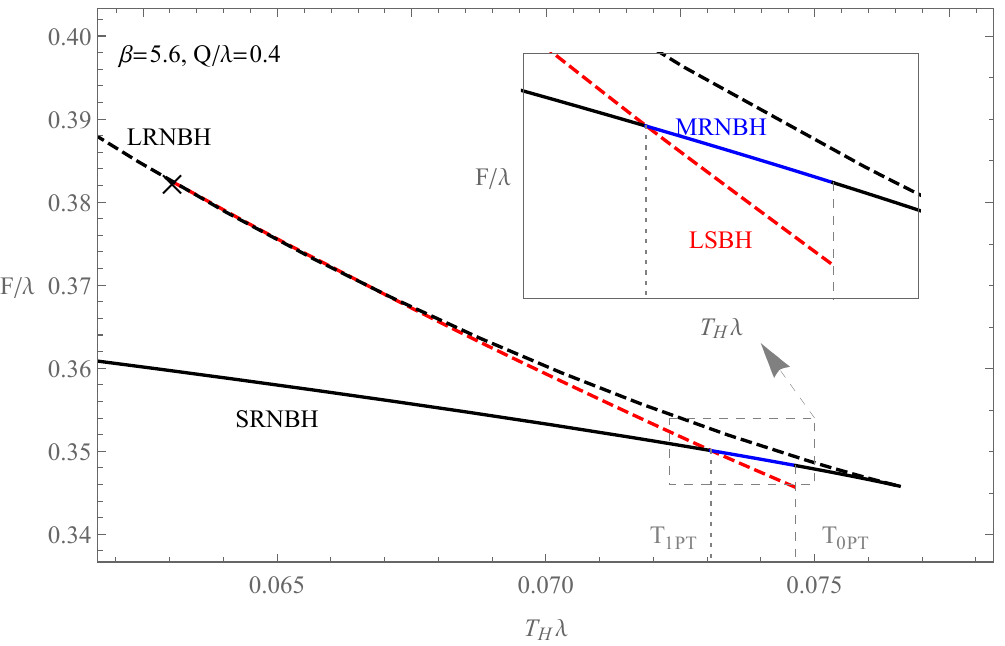}
    \caption{Thermodynamic behavior of the solutions for $\beta=5.6$. The curves and symbols follow the same conventions as in Fig.~\ref{thermanalysisbeta6}.}
    \label{thermanalysisbeta5p6}
\end{figure}

The thermal behavior of the system for $\beta=5.6$ is illustrated in Fig.~\ref{thermanalysisbeta5p6}. In this case, the charged scalarized branches shrink appreciably and become a Schwarzschild-like single branch that remain locally unstable throughout their entire range of existence, with no associated Davies point. In this regime, the coupling between the scalar field and the Gauss-Bonnet invariant is sufficiently strong that the effect of the electromagnetic field becomes negligible. Consequently, the SRNBH configuration constitutes the thermodynamically stable preferred phase of the system, only divided by an intermediate locally unstable LSBH that dominates the partition function for a small temperature range. These generate two phase transitions: a zeroth-order transition in $T_{0\mathrm{PT}}$ and a first-order transition in $T_{1\mathrm{PT}}$.

A black hole residing on the SRNBH branch for $T > T_{0\mathrm{PT}}$ will cool as it evaporates, undergoing a zeroth-order phase transition to the LSBH. The LSBH then continues to evaporate until it returns, through the same phase transition, to the stable SRNBH branch. For $T < T_{1\mathrm{PT}}$, an SRNBH follows an analogous evolution under mass accretion. If a black hole lies within the locally unstable region of the scalarized branch, its evolution depends on the direction of mass change. Under accretion, it moves leftward in the diagram and undergoes a first-order phase transition to the SRNBH phase. Under evaporation, it moves rightward and instead undergoes a zeroth-order phase transition to the SRNBH phase. In all cases, re-entrant phase transitions do not occur without external contrivance. That is, under monotonic variation of the temperature, the system does not return to its initial thermodynamic phase through a continuous natural process.
 
For sufficiently strong coupling, the charged scalarized branch shrinks significantly and ceases to support thermodynamically dominant hairy configurations. In this regime, the SRNBH branch remains the only thermodynamically stable phase and no phase transitions occur. As noted above, this reflects the strong effect of the coupling as the electromagnetic field becomes negligible in this limit, and charged scalarized black holes closely resemble their neutral counterparts in a strongly restricted domain of existence.

\begin{figure}[h]  
\centering
    \includegraphics[width=0.45\textwidth]{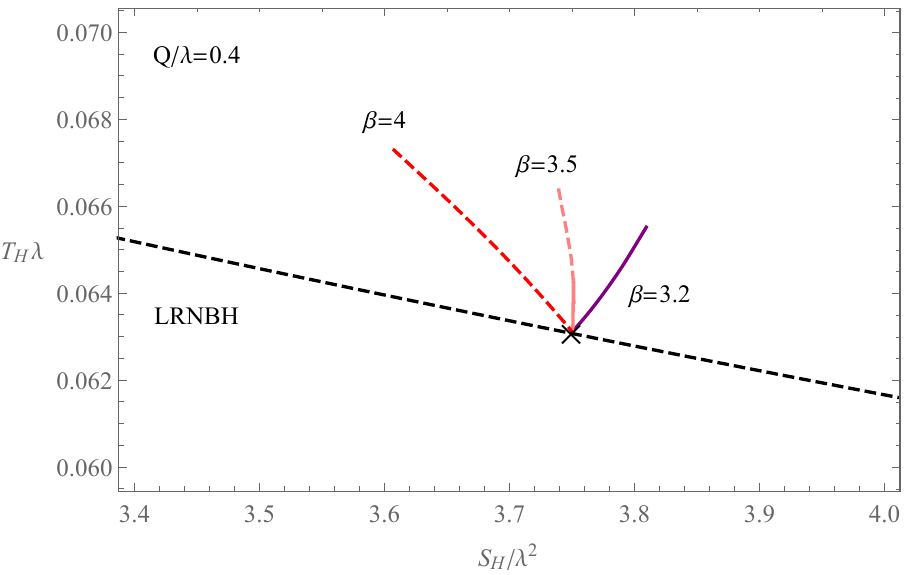} 
    \caption{Equation of state given by the Hawking temperature $T_H$ as a function of the entropy $S_H$, shown for three values of the coupling constant, $\beta = 4, 3.5, 3.2$, in the strong-coupling regime. The curves and symbols follow the same conventions as in Fig.~\ref{thermanalysisbeta48}.}
    \label{equationofstateTvsSQ04strongcouplingregime}
\end{figure}

In Fig.~\ref{equationofstateTvsSQ04strongcouplingregime}, we display the equation of state $T(S)$ for the charged case at different values of the coupling constant $\beta$. For $\beta = 3.5$, the charged scalarized branch is locally stable near the bifurcation point and exhibits a turning point beyond which it becomes locally unstable. For $\beta = 3.2$, the scalarized branch remains locally stable throughout its entire domain of existence. This behavior is consistent with that found for neutral scalarized black holes in the strong-coupling regime \cite{Herdeiro:2026sur}, confirming that, in this limit, charged solutions closely track the properties of their neutral counterparts.

Nevertheless, even when locally stable, the charged scalarized configurations are never thermodynamically favored, as their Helmholtz free energy consistently exceeds that of the SRNBH branch. Consequently, they do not dominate the partition function but appear only as thermodynamically disfavored solutions, as in the case of $\beta = 4.0$, or as metastable states, as in the case of $\beta = 3.2$.

\section{Thermodynamic phase structure}\label{phase structure}

The analysis of Sec.~\ref{Thermodynamical analysis} shows how the free energy determines the thermodynamic phases available to the system as a function of the coupling strength and temperature, and, in particular, how the phase transition temperatures depend on the coupling parameter $\beta$.

\begin{figure}[h]  
\centering
    \includegraphics[width=0.45\textwidth]{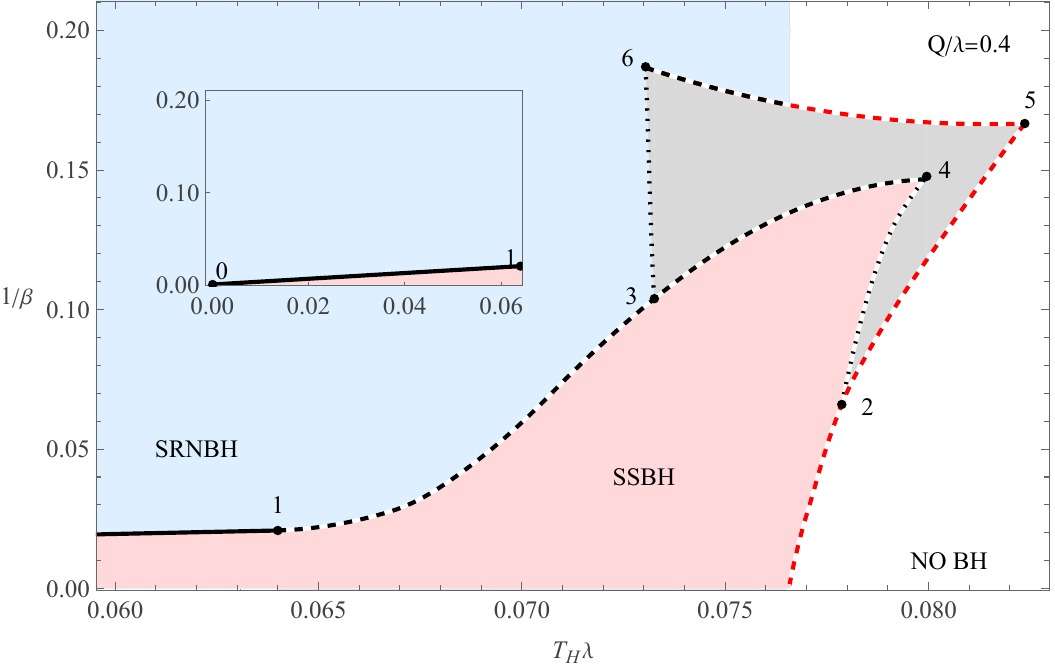} 
  \caption{Phase structure in the $1/\beta$-$T$ plane over a range of coupling strengths for which scalarized black holes and phase transitions occur. Here $Q/\lambda = 0.4$. The shaded region indicates the domain where the black hole solutions are locally thermodynamically unstable under small thermal fluctuations. The black solid, dotted, and dashed curves represent second-, first-, and zeroth-order phase transitions, respectively. The red dashed curve corresponds to the maximal temperature of the scalarized branches. The labeled black dots mark threshold values of the coupling $1/\beta$ at which the qualitative structure of the phase diagram changes; they are ordered by increasing $1/\beta$ and should be read as points on the vertical axis, not as endpoints of the transition curves.}
  \label{phasestructureBETAvsTchargedQ04}
\end{figure}

The resulting phase diagram is summarized in Fig.~\ref{phasestructureBETAvsTchargedQ04}, presented in the $1/\beta$-$T_H$ plane, where $1/\beta$ serves as a measure of the scalar-curvature coupling strength. In this parametrization, weak coupling corresponds to small values of $1/\beta$, while strong coupling corresponds to larger ones.

Two thermodynamically stable phases can be identified, corresponding to the SRNBH and SSBH branches, together with a region of large scalarized black holes that remain locally thermodynamically unstable (shown as a shaded area). Depending on the value of the coupling constant $\beta$, the phase structure can be organized into three qualitatively distinct regimes, each admitting a further sub-structure. These regimes are delimited by the labeled points of Fig.~\ref{phasestructureBETAvsTchargedQ04}, which mark successive threshold values of the coupling $1/\beta$ on the vertical axis, ordered by increasing coupling strength. Accordingly, an expression such as ``between points $i$ and $j$'' always denotes the interval of coupling strengths delimited by the corresponding values of $1/\beta$, and never a segment joining two points in the $(T_H,1/\beta)$ plane. Within each such interval the phase structure is qualitatively unchanged, while crossing a point marks a qualitative change.

The weak-coupling regime (between points $0$ and $2$) admits two sub-regimes, both characterized by the presence of two equilibrium phases connected by genuine\footnote{We reserve the terms \emph{genuine} or \emph{bona fide} for transitions between equilibrium phases, thereby distinguishing them from transitions involving thermodynamically unstable configurations.} thermodynamic phase transitions. For couplings between points $0$ and $1$, the two stable phases are connected by a second-order phase transition (solid curve) whose transition temperature $T_{2\mathrm{PT}}$ gradually increases with the coupling strength (see Fig.~\ref{thermanalysisbeta48}). For couplings between points $1$ and $2$, the scalarized branch progressively shrinks and the transition becomes zeroth order (dashed curve), with a finite jump in the free energy separating the two stable phases; the corresponding transition temperature $T_{0\mathrm{PT}}$ increases with $1/\beta$ (see Fig.~\ref{thermanalysisbeta24}). 

The intermediate-coupling regime (between points $2$ and $5$) admits three sub-regimes, all characterized by the presence of locally unstable regions in the phase structure and the growing metastability of the SSBH configurations. For couplings between points $2$ and $3$, a characteristic fish-like structure emerges in the Helmholtz free energy, giving rise to a shaded domain in which large scalarized black holes possess lower free energy than the small scalarized ones in the vicinity of their Davies point, rendering SSBH partially metastable (see Fig.~\ref{thermanalysisbeta10}). For couplings between points $3$ and $4$, the scalarized branch shrinks further, leading to the emergence of an additional unstable region in which large scalarized configurations have lower free energy than the SRNBH branch (see Fig.~\ref{thermanalysisbeta8}). For couplings between points $4$ and $5$, the SSBH branch becomes entirely metastable and the LSBH configurations dominate the high-temperature region of the phase diagram (see Fig.~\ref{thermanalysisbeta6}).

The strong-coupling regime (beyond point $5$) is characterized by Schwarzschild-like scalarized branches that resemble their neutral counterparts. As the coupling increases, these branches shrink (see Fig.~\ref{thermanalysisbeta5p6}), until they disappear from the phase diagram at point $6$, beyond which the SRNBH phase constitutes the sole thermodynamically preferred configuration at all temperatures compatible with extremality (see Fig.~\ref{equationofstateTvsSQ04strongcouplingregime}).

It is worth emphasizing that both transition lines confined to the weak-coupling regime, the second-order and zeroth-order ones, represent genuine phase transitions between two locally stable phases, the former being a continuous transition and the latter characterized by a finite jump in the free energy. Unlike these, the additional transitions arising in the intermediate- and strong-coupling regimes involve locally unstable configurations and, therefore, do not constitute phase transitions between equilibrium phases. The special points appearing in the phase diagram are also not thermodynamic critical points. They signal structural changes in the free-energy landscape, such as the termination of transition lines, the onset of multi-branch behavior, or the contraction of the scalarized branch to the point of thermodynamic irrelevance, and therefore mark qualitative shifts in the global phase structure rather than the emergence of critical behavior.

The shaded region corresponds to a domain in which the black hole solutions are locally thermodynamically unstable and therefore do not represent physically realizable equilibrium configurations. From the perspective of black hole thermodynamics, this region can be regarded as a \emph{terra incognita} of the phase diagram \cite{Chamblin:1999hg}, raising the question of which equilibrium configurations, if any, may reside there. One possible interpretation is that the system transitions to states of lower free energy through dynamical mechanisms beyond curvature-induced scalarization. Recent work suggests that a strong violation of black hole uniqueness may arise from the coexistence of multiple scalarization mechanisms, leading to several distinct branches of solutions beyond the trivial one \cite{Eichhorn:2026qaw}. Exploring such possibilities lies beyond the scope of the present work, and we leave a detailed investigation of this scenario for future work.

\section{Conclusions}\label{conclusions}

Asymptotically flat black holes are typically characterized by the absence of thermodynamic phase transitions in the canonical ensemble. In this work, we have shown that this need not be the case when scalarization is at play. Working within the canonical ensemble and employing the Euclidean approach to construct the corresponding thermodynamic potentials, we have demonstrated that Reissner-Nordstr\"om and curvature-induced scalarized black holes in Einstein-Maxwell-scalar-Gauss-Bonnet theory exhibit a rich phase structure, without invoking external confining mechanisms or an extended thermodynamic formalism.

The resulting thermodynamic phase structure is highly sensitive to the coupling constant $\beta$ and is naturally organized into three qualitatively distinct regimes. In the weak-coupling regime, the charged scalarized branch inherits the local stability structure of the RN solution, exhibiting a Davies point that separates locally stable from locally unstable configurations. The Helmholtz free energy reveals two stable phases, the SRNBH and the SSBH, connected by a second-order phase transition that coincides with the second bifurcation point at which the scalarized branch reconnects with the RN branch, signaling the spontaneous loss of scalar hair. As the coupling strength increases, the scalarized branch departs from its RN counterpart and progressively shrinks; in turn, the transition becomes zeroth order, with a finite jump in the free energy separating the two stable phases. In the intermediate-coupling regime, the free energy develops a characteristic fish-like structure in the vicinity of the Davies point of the scalarized branch, from which a first-order transition between the SSBH and LSBH phases emerges. In this process, locally stable scalarized configurations near the Davies point become metastable. With increasing coupling, the LSBH branch becomes thermodynamically favored over the SRNBH phase as well, generating an additional first-order transition and giving rise to up to three transitions at distinct temperatures. Toward the end of this regime, the SSBH branch becomes entirely metastable and the LSBH configurations dominate the high-temperature region of the phase diagram. In the strong-coupling limit, the scalarized branch reduces to a Schwarzschild-like solution and eventually shrinks to the point of becoming thermodynamically irrelevant, leaving the SRNBH phase as the sole thermodynamically preferred configuration.

The emergence of this nontrivial thermodynamic behavior hinges on two key ingredients: the existence of a region of local thermodynamic stability for small Reissner-Nordstr\"om black holes, and a scalar field nonminimally coupled to the Gauss-Bonnet invariant. Neither ingredient alone suffices, and it is only their interplay that elevates free-energy crossings to bona fide phase transitions between equilibrium phases. To justify this claim, it is instructive to consider two limiting cases.

In the neutral case, the Gauss-Bonnet coupling is present but no region of local stability exists for the Schwarzschild black hole, so the free-energy comparison between the trivial and scalarized branches does not involve two locally stable phases \cite{Herdeiro:2026sur}. In the weak-coupling regime, the neutral scalarized branch mimics the local instability of the Schwarzschild branch, and a second-order phase transition occurs precisely at the first bifurcation point, i.e. at the onset of scalarization. In the charged case studied here, the relevant second-order transition is tied to the second bifurcation point rather than the first, reflecting the coexistence of two equilibrium phases at the transition, a feature with no analog in the neutral case. 

The second limiting case, in which the scalar field couples to the Maxwell term rather than to the Gauss-Bonnet invariant, yields an equally distinct thermodynamic picture. In Einstein-Maxwell-scalar theory \cite{Santos:2022vet}, matter-induced scalarization produces charged scalarized black holes that, despite bifurcating from locally stable RN branches, are thermodynamically unstable in the canonical ensemble, leaving no room for the rich phase structure reported here. This contrast highlights that curvature-induced and matter-induced scalarization mechanisms play fundamentally different thermodynamic roles, producing qualitatively distinct phase structures in the canonical ensemble, reflecting their different dynamical behavior.

Several directions remain open. A natural question is whether the connection between the second bifurcation point and a second-order thermodynamic transition is a universal feature of curvature-induced scalarization in the charged case or whether it depends on specific properties of the coupling function, a correlation that would link a genuine thermodynamic phenomenon to a purely dynamical one. It would also be interesting to extend the present analysis to frameworks in which the scalar field couples nonminimally to both the Gauss-Bonnet invariant and the Maxwell field \cite{Belkhadria:2025lev}, in order to assess the thermodynamic interplay between curvature- and matter-induced scalarization. A further natural extension is to spin-induced scalarization, which would probe the analogous competition between rotating scalarized and Kerr black holes. More broadly, understanding how multiple scalarization mechanisms may coexist and interact \cite{Eichhorn:2026qaw} offers a promising route toward clarifying the locally unstable \emph{terra incognita} regions identified in the phase diagram and toward a more complete picture of thermodynamic phase structures in asymptotically flat black hole spacetimes.

Finally, our analysis has addressed thermodynamic stability alone, and it is natural to ask how the phase structure reported here relates to the dynamical stability of the corresponding configurations. In the neutral case, the scalarized branches obtained with the Gaussian coupling \eqref{gaussiancoupling} have been studied under radial \cite{BlazquezSalcedo:2018jnn, BlazquezSalcedo:2022omw}, axial \cite{BlazquezSalcedo:2020rhf} and polar \cite{BlazquezSalcedo:2020caw} perturbations, revealing both dynamically stable and unstable sectors, as well as a loss of hyperbolicity of the perturbation equations below a critical horizon size. We expect analogous features to persist in the presence of an electric charge. Such an analysis would determine which of the locally stable phases identified here are also dynamically realizable, all the more so since the two notions of stability are logically independent: in the neutral case, the radially stable scalarized black holes possess negative specific heat. A complete characterization of these thermal states will therefore require the thermodynamic picture developed here to be combined with a full perturbative analysis of the charged scalarized branches.

\section*{Acknowledgments}
C.E. is supported by Universidad Central de Chile through Proyecto Puente de Investigación grant CIPPTE202605. C.E. also thanks the National Technical University of Athens for its hospitality during the initial stage of this work. The research of S.K. is supported by the FONDECYT Postdoctorado No.~3250501, a program of the Agencia Nacional de Investigación y Desarrollo (ANID), Chile. C.E. and S.K. thank R. B. Mann for valuable feedback at the final stage of this work, during the Third Pacific Gravity School: New Horizons in Gravitational Physics.

\providecommand{\href}[2]{#2}\begingroup\raggedright
\endgroup

\end{document}